\documentclass[twoside,english]{iopart}
\usepackage[T1]{fontenc}
\usepackage[latin9]{inputenc}
\usepackage{geometry}
\usepackage{units}
\usepackage{amstext}
\usepackage{graphicx}
\usepackage{esint}

\makeatletter
\usepackage{iopams}
\usepackage{setstack}

\usepackage[numbers, sort&compress]{natbib}

\newcommand{\eqref}[1]{(\ref{#1})}

\makeatother

\usepackage{babel}
\begin{document}

\title[Vector potential and physical implications]{An educational path for the magnetic vector potential and its physical
implications}

\author{{\Large{S Barbieri$^{1}$, M Cavinato$^{2}$ and M Giliberti$^{2}$
}}}

\address{{\Large{$^{1}$}}{\large{Università degli Studi di Palermo, Italy}}}

\address{{\Large{$^{2}$}}{\large{Dipartimento di Fisica, Università degli
Studi di Milano, Italy}}}

\ead{marco.giliberti@unimi.it}
\begin{abstract}
We present an educational path on the magnetic vector potential $\mathbf{A}$
addressed to undergraduate students and to pre-service physics teachers.
Starting from the generalized Ampère-Laplace law, in the framework
of a slowly varying time-dependent field approximation, the magnetic
vector potential is written in terms of its empirical referent, i.
e. the conduction current. Therefore, once the currents are known,
our approach allows a clear and univocal physical determination of
$\mathbf{A}$, overcoming the mathematical indeterminacy due to the
gauge transformations. We have no need to fix a gauge, since for slowly
varying time-dependent electric and magnetic fields, the ``natural''
gauge for $\mathbf{A}$ is the Coulomb one. We stress the difference
between our approach and those usually presented in the literature.
Finally, a physical interpretation of the magnetic vector potential
is discussed and some examples of calculation of $\mathbf{A}$ are
analysed.
\end{abstract}

\noindent{\it Keywords\/}: {Vector potential, Electromagnetic field, Education}

\pacs{03.50.De, 01.55.+b}

\ams{}

\submitto{European Journal of Physics}

\maketitle

\section{Introduction}

The magnetic vector potential $\mathbf{A}$ is very useful in many
physical situations. Besides its obvious relevance in standard quantization
of the electromagnetic field and in electromagnetic gauge theories,
$\mathbf{A}$ is fundamental in understanding both some classical
phenomena (i. e. Maxwell-Lodge effect \cite{Iencinella,Rousseaux})
and some quantum physics phenomena (i. e. Aharonov-Bohm effect \cite{Aharonov}
and Mercereau effect \cite{Mercereau}). Moreover it gives the possibility
of introducing superconductivity in a simple and meaningful way, at
least within a phenomenological approach \cite{London,Pippard,Barbieri}.
Nevertheless, from a teaching point of view, in many introductory
textbooks on electromagnetism, as also in undergraduate up to graduate
lectures, the magnetic vector potential is generally presented only
as a useful mathematical tool, disregarding its physical meaning.
Even if many papers can be found in literature that clarify that vector
potential does indeed have a precise physical meaning \cite{Iencinella,Giuliani,Konopinski,Semon},
nonetheless a clear educational path on vector potential is, to the
best of our knowledge, still missing. Therefore, as a part of a PhD
research in physics education, we have developed an approach to magnetic
vector potential that is now being tested in three different experimentations\cite{Cavinato}.
The first one is addressed to third year college students in mathematics
and is being carried out in a basic course on electromagnetism. The
second one is addressed to pre-service physics teachers and is being
carried out in a course on electromagnetic induction education. The
last one is a laboratory course addressed to graduate students in
mathematics. In what follows we present the general framework of our
educational path on the magnetic vector potential. Our intention is
to show a particularly meaningful way to introduce the vector potential
for slowly varying time-dependent fields in terms of an empirical
referent, i.e. the conduction current density. The \textquotedblleft{}natural\textquotedblright{}
gauge condition is then discussed and thoughts on the physical meaning
of $\mathbf{A}$ with examples are finally offered. The results obtained
in the previous mentioned experimentations will be, instead, discussed
in a forthcoming paper.

\section{Our educational path on the magnetic vector potential}

Maxwell in his book ``A treatise on electricity and magnetism''
introduced the notion of magnetic vector potential through an integral
relation. Following Maxwell, but with modern symbology, we can say
that the magnetic vector potential\textbf{ A} is a vector\textbf{
}such that the flux of the magnetic field \textbf{B} through any surface
$\Sigma$ is equal to the circulation of \textbf{A} around the boundary
$\partial\Sigma$ of $\Sigma$ \cite{Maxwell}, that is:

\begin{equation}
\int_{\Sigma}\mathbf{B}\cdot\mathbf{n}d\Sigma=\oint_{\partial\Sigma}\mathbf{A}\cdot d\mathbf{s}.\label{eq:circuitazione}
\end{equation}
On the contrary, textbooks usually introduce the magnetic vector potential
via a differential equation that is the local form of eq. (\ref{eq:circuitazione});\textbf{
A} is defined as the vector such that:

\begin{equation}
\mathbf{B}=\nabla\times\mathbf{A}.\label{B =00003D rot A}
\end{equation}
As it is well known, neither equation (1) nor equation (2) give an
explicit definition of\textbf{ A}. In fact, for a specified magnetic
field \textbf{B}, there are many possible solutions for \textbf{A
}of both equation (1) and (2); that is,\textbf{ A} is not univocally
defined by\textbf{ B}. This fact is at the basis of the so called
gauge invariance that will be discussed later, and is one of the main
reasons of many difficulties in understanding the physical meaning
of the magnetic vector potential \cite{Giliberti}. In the following
we develop a path on the magnetic vector potential that we believe
is much more meaningful than the traditional one; it is addressed
to both undergraduate students and secondary school teachers. 

Given a general distribution of conduction current density\textbf{
J} and taking also into account the displacement current density,
the magnetic field\textbf{ B} at position \textbf{r} and time \textit{t},
in vacuum, is given by the generalized Ampère-Laplace law: 
\begin{equation}
\mathbf{B}\left(\mathbf{r},t\right)=\frac{\mu_{0}}{4\pi}\int_{V'}\frac{\left[\mathbf{J}\left(\mathbf{r}',t'\right)+\varepsilon_{0}\frac{\partial\mathbf{E}\left(\mathbf{r}',t'\right)}{\partial t}\right]\times\Delta\mathbf{r}}{\left(\Delta r\right)^{3}}dV',\label{Ampere Laplace}
\end{equation}
where $V'$ is the volume containing the currents and
\begin{equation}
\Delta\mathbf{r}\equiv\mathbf{r}-\mathbf{r}',\:\Delta r\equiv\left|\Delta\mathbf{r}\right|,\: t'\equiv t-\frac{\Delta r}{c},\label{eq:notazione}
\end{equation}
where $t'$ is the retarded time. If we now adopt the quasi-static
approximation, that is if we consider only fields that are slowly
varying in time, we can neglect all the time derivative multiplied
by $\nicefrac{1}{c}$ (but not time dependent terms alone). Therefore
the contribution of the displacement currents in equation (\ref{Ampere Laplace})
can be disregarded, thanks to the presence of the constant $\varepsilon_{0}\mu_{0}$$=1/c^{2}$
that multiplies the time derivative of $\mathbf{E}$. Moreover, also
the retarded time $t'$ of equation (\ref{eq:notazione}) can be considered
equal to $t$. So we are left with 
\begin{equation}
\mathbf{B}\left(\mathbf{r},t\right)=\frac{\mu_{0}}{4\pi}\int_{V'}\frac{\mathbf{J}\left(\mathbf{r}',t\right)\times\Delta\mathbf{r}}{\left(\Delta r\right)^{3}}dV'.\label{eq:B quasi statico}
\end{equation}
Observing that:
\begin{equation}
\nabla\left(\frac{1}{\Delta r}\right)=-\frac{\Delta\mathbf{r}}{\left(\Delta r\right)^{3}},\label{eq:gradiente uno su erre}
\end{equation}
and commutating the factors of the vector product in the integrand
of equation (\ref{eq:B quasi statico}), we can write:
\begin{equation}
\mathbf{B}\left(\mathbf{r},t\right)=\frac{\mu_{0}}{4\pi}\int_{V'}\nabla\left(\frac{1}{\Delta r}\right)\times\mathbf{J}\left(\mathbf{r}',t\right)dV'.\label{eq:B quasi statico uno}
\end{equation}
Keeping in mind that if\textit{ f }is a scalar field while \textbf{v}
is a vector field one has the identity:
\begin{equation}
\nabla\times\left(f\mathbf{v}\right)=\nabla f\times\mathbf{v}+f\nabla\times\mathbf{v}\label{eq:regola rotore}
\end{equation}
and using the fact that $\nabla\times\mathbf{J}(\mathbf{r}',t)=0$
because \textbf{J} depends on primed variables while the curl is done
with respect to unprimed ones, we obtain:
\begin{equation}
\mathbf{B}\left(\mathbf{r},t\right)=\nabla\times\left(\frac{\mu_{0}}{4\pi}\int_{V'}\frac{\mathbf{J}\left(\mathbf{r}',t\right)}{\Delta r}dV'\right).\label{eq:B quasi statico finale}
\end{equation}
Equation (\ref{eq:B quasi statico finale}) clearly shows that we
can define a vector:
\begin{equation}
\mathbf{A}\left(\mathbf{r},t\right)\equiv\frac{\mu_{0}}{4\pi}\int_{V'}\frac{\mathbf{J}\left(\mathbf{r}',t\right)}{\Delta r}dV'\label{eq:A in funzione di J}
\end{equation}
such that
\begin{equation}
\mathbf{B}\left(\mathbf{r},t\right)=\nabla\times\mathbf{A}\left(\mathbf{r},t\right).\label{eq:B rotore di A}
\end{equation}

Equation (\ref{eq:A in funzione di J}) proves a clear analogy between
magnetic vector potential and electric scalar potential 
\begin{equation}
\mathbf{\varphi}\left(\mathbf{r},t\right)=\frac{1}{4\pi\varepsilon_{0}}\int_{V'}\frac{\mathbf{\rho}\left(\mathbf{r}',t\right)}{\Delta r}dV',\label{eq:electric potential}
\end{equation}
where $\mathbf{\rho}\left(\mathbf{r}',t\right)$ is the charge density
at point $\mathbf{r}'$ and time $t$. With the definition given by
equation (\ref{eq:A in funzione di J}) the vector potential is a
precise function of the current density. Therefore (in our slow varying
field approximation) once the currents are known,\textbf{ A }is univocally
determined. Instead, if we define the vector potential (as it is usually
done) through equation (\ref{B =00003D rot A}), it is not surprising
that it is not univocally determined. In fact, Helmholtz theorem (and
its generalized version \cite{Heras}) states that a quasi-static
vector field vanishing at infinity more quickly than 1/\textit{r},
is completely determined once both its curl \textit{and} its divergence
are known. Therefore an additional condition (the so-called gauge
condition) is clearly needed. This is generally done by arbitrary
fixing the divergence of \textbf{A}. On the contrary, with our definition
(equation (\ref{eq:A in funzione di J})), we have no need to fix
a gauge. Nevertheless to better understand the link between our explicit
definition (equation (\ref{eq:A in funzione di J})) and the usual
one (\ref{B =00003D rot A}), it is interesting to determine the gauge
we are implicitly using. This can be done by directly calculating
$\nabla\cdot\mathbf{A}$. 

With the same symbology of equation (\ref{eq:regola rotore}) we have
the following vector identity: 
\begin{equation}
\nabla\cdot\left(f\mathbf{v}\right)=\nabla f\cdot\mathbf{v}+f\nabla\cdot\mathbf{v},\label{eq:divergenza}
\end{equation}
therefore, from eq. (\ref{eq:A in funzione di J}) we obtain:
\begin{equation}
\mathbf{\nabla\cdot A}\left(\mathbf{r},t\right)=\frac{\mu_{0}}{4\pi}\int_{V'}\nabla\left(\frac{1}{\Delta r}\right)\cdot\mathbf{J}\left(\mathbf{r}',t\right)dV'+\frac{\mu_{0}}{4\pi}\int_{V'}\frac{1}{\Delta r}\nabla\cdot\mathbf{J}\left(\mathbf{r}',t\right)dV'\label{eq:divergenza A primo}
\end{equation}
\[
\,\,\,\,\,\,\,\,\,\,\,\,\,\,\,\,\,\,\,\,\,\,\,\,\,\,\,\,\,\,=-\frac{\mu_{0}}{4\pi}\int_{V'}\nabla^{'}\left(\frac{1}{\Delta r}\right)\cdot\mathbf{J}\left(\mathbf{r}',t\right)dV',
\]
where the nabla symbol $\nabla^{'}$operates with respect to primed
variables. We note that $\nabla\cdot\mathbf{J}\left(\mathbf{r}',t\right)=0$
because\textbf{ J} depends only on primed variables while the divergence
is done with respect to unprimed ones and $\nabla\left(\frac{1}{\Delta r}\right)=-\nabla'\left(\frac{1}{\Delta r}\right)$.
Moreover (again keeping in mind equation (\ref{eq:divergenza})) the
integrand of the last term in equation (\ref{eq:divergenza A primo})
can be written as follows: 
\begin{equation}
\nabla'\left(\frac{1}{\Delta r}\right)\cdot\mathbf{J}\left(\mathbf{r}',t\right)=\nabla^{'}\cdot\left[\frac{\mathbf{J}\left(\mathbf{r}',t\right)}{\Delta r}\right]-\frac{1}{\Delta r}\nabla'\cdot\mathbf{J}\left(\mathbf{r}',t\right)\label{eq: integrando della 14}
\end{equation}
\[
\,\,\,\,\,\,\,\,\,\,\,\,\,\,\,\,\,\,\,\,\,\,\,\,\,\,\,\,\,\,\,\,\,\,\,\,\,\,\,\,\,\,\,\,\,\,\,\,\,\,=\nabla^{'}\cdot\left[\frac{\mathbf{J}\left(\mathbf{r}',t\right)}{\Delta r}\right]+\frac{1}{\Delta r}\frac{\partial\mathbf{\rho}\left(\mathbf{r}',t\right)}{\partial t},
\]
where, in the last equality, we have used the continuity equation:
\begin{equation}
\nabla\cdot\mathbf{J}\left(\mathbf{r},t\right)+\frac{\partial\mathbf{\rho}\left(\mathbf{r},t\right)}{\partial t}=0.\label{eq:continuit=0000E0}
\end{equation}
Therefore we have:
\begin{equation}
\mathbf{\nabla\cdot A}\left(\mathbf{r},t\right)=-\frac{\mu_{0}}{4\pi}\int_{V'}\nabla^{'}\cdot\left[\frac{\mathbf{J}\left(\mathbf{r}',t\right)}{\Delta r}\right]dV'-\frac{\mu_{0}}{4\pi}\int_{V'}\frac{1}{\Delta r}\frac{\partial\mathbf{\rho}\left(\mathbf{r}',t\right)}{\partial t}dV'.\label{eq:divAsecondo}
\end{equation}
The first integral in equation (\ref{eq:divAsecondo}) is zero thanks
to the divergence theorem. In fact
\begin{equation}
-\frac{\mu_{0}}{4\pi}\int_{V'}\nabla^{'}\cdot\left[\frac{\mathbf{J}\left(\mathbf{r}',t\right)}{\Delta r}\right]dV'=-\frac{\mu_{0}}{4\pi}\int_{\Sigma'}\frac{\mathbf{J}\left(\mathbf{r}',t\right)}{\Delta r}\cdot\mathbf{n}d\Sigma',\label{eq:teodivI}
\end{equation}
where $\Sigma'\equiv\partial V'$ is the boundary of the volume $V'$
and $\mathbf{n}$ is the outer normal to $\Sigma'$. Since $V'$ must
contain at each time all the currents that generate \textbf{A}, it
can be taken so large that\textbf{ J} can be considered zero upon
$\Sigma'$ and therefore the right hand side integral in equation
(\ref{eq:teodivI}) vanishes. For the second integral in the right
hand side of equation (\ref{eq:divAsecondo}), since $V'$ is time-independent,
we get:
\begin{equation}
\frac{\mu_{0}}{4\pi}\int_{V'}\frac{1}{\Delta r}\frac{\partial\mathbf{\rho}\left(\mathbf{r}',t\right)}{\partial t}dV'=\frac{\mu_{0}}{4\pi}\frac{\partial}{\partial t}\int_{V'}\frac{\mathbf{\rho}\left(\mathbf{r'},t\right)}{\Delta r}dV'=\frac{\mu_{0}}{4\pi}\frac{\partial}{\partial t}\left[4\pi\varepsilon_{0}\mathbf{\varphi}\left(\mathbf{r},t\right)\right],\label{eq:integrale II}
\end{equation}
where $\mathbf{\varphi}\left(\mathbf{r},t\right)$ is the electric
scalar potential given by equation (\ref{eq:electric potential}).
From equations (\ref{eq:divAsecondo}) - (\ref{eq:integrale II})
we finally obtain:
\begin{equation}
\mathbf{\nabla\cdot A}\left(\mathbf{r},t\right)=-\varepsilon_{0}\mu_{0}\frac{\partial}{\partial t}\mathbf{\varphi}\left(\mathbf{r},t\right)=-\frac{1}{c^{2}}\frac{\partial}{\partial t}\mathbf{\varphi}\left(\mathbf{r},t\right).\label{eq:gaugedi lorenz}
\end{equation}

Equation (\ref{eq:gaugedi lorenz}) is the well-known Lorenz gauge.
In the quasi-static approximation we are adopting in this paper, the
right hand term of equation (\ref{eq:gaugedi lorenz}) can be considered
zero, and therefore we are left in the so-called Coulomb gauge: 
\begin{equation}
\mathbf{\nabla\cdot A}\left(\mathbf{r},t\right)=0.\label{eq:gaugediCoulomb}
\end{equation}
The most common attitude is to define\textbf{ A} from equation (\ref{eq:B rotore di A}),
to arbitrary fix $\mathbf{\nabla\cdot A}\left(\mathbf{r},t\right)=0$
from the beginning so that equation (\ref{eq:A in funzione di J})
is obtained as a final result. On the contrary, we have followed an
inverse path in which we have been naturally led to define a magnetic
vector potential in terms of the current density (that can therefore
be seen as the source of the potential) as we have done in equation
(\ref{eq:A in funzione di J}). Only as a consequence of this definition
we found that the previously defined magnetic vector potential is
given in the Coulomb gauge, that therefore can be seen as the \textquotedblleft{}natural\textquotedblright{}
gauge for slowly varying fields. 

As it is well known the relations which give a link among the fields
$\mathbf{E}\left(\mathbf{r},t\right)$, $\mathbf{B}\left(\mathbf{r},t\right)$
and the potentials are: 
\begin{equation}
\mathbf{E}\left(\mathbf{r},t\right)=-\frac{\partial}{\partial t}\mathbf{A}\left(\mathbf{r},t\right)-\nabla\mathbf{\varphi}\left(\mathbf{r},t\right)\label{eq:E coi potenziali}
\end{equation}
 and equation (\ref{eq:B rotore di A}). In the general case, when
the following trasformations (called gauge transformations) are performed:
\begin{equation}
\varphi\rightarrow\varphi'=\varphi-\frac{\partial\Lambda}{\partial t}\label{eq:tras scal}
\end{equation}
\begin{equation}
\mathbf{A}\rightarrow\mathbf{A'}=\mathbf{A}+\nabla\Lambda,\label{eq:tras vet}
\end{equation}
where $\Lambda$ is a scalar function, the electric and the magnetic
fields remain unchanged. Here we want to stress that our choice to
define the vector potential in terms of the convective current (equation(\ref{eq:A in funzione di J}))
is not equivalent to the choice of the Coulomb gauge, where \textbf{A}
is determined only up to the gradient of a harmonic function, as can
be seen from equation (\ref{eq:tras vet}).

\section{The physical meaning of the magnetic vector potential}

The fact that the scalar potential is determined only up to the time
derivative of a scalar function does not prevent us from giving it
a physical meaning when the electric field is slowly varying in time.
In a similar (but \textquotedblleft{}dual\textquotedblright{}) way,
the fact that the vector potential is determined only up to the space
derivative of a scalar function does not prevent us from giving it
a physical meaning when the magnetic field is slowly varying in time. 

When the vector potential \textbf{A} in equation (\ref{eq:E coi potenziali})
is time independent, we can give a physical meaning to the scalar
function $\varphi$. In fact, in this condition, we can define the
potential energy $\mathbf{\mathit{U}}\left(\mathbf{r},t\right)$ of
a point charge \textit{q }set at position \textbf{r} and time \textit{t
}as the work (independent of the chosen path) necessary to move the
charge \textit{q }from inf

nity, where the electric field is zero, to the point \textbf{r}, against
the forces of the electric field; that is:
\begin{equation}
\mathbf{\mathit{U}}\left(\mathbf{r},t\right)=-\int_{\infty}^{\mathbf{r}}q\mathbf{E}\left(\mathbf{r'},t\right)\cdot d\mathbf{r'}.\label{eq:en pot scal}
\end{equation}
The electric scalar potential can therefore be written as: 
\begin{equation}
\mathbf{\mathit{\varphi}}\left(\mathbf{r},t\right)=-\int_{\infty}^{\mathbf{r}}\mathbf{E}\left(\mathbf{r'},t\right)\cdot d\mathbf{r'}.\label{eq:electric pot}
\end{equation}
It has the clear physical meaning of potential energy per unit charge
and can be identified with the function $\varphi$ of equation (\ref{eq:E coi potenziali}).
We note that the integrals of equations (\ref{eq:en pot scal}) and
(\ref{eq:electric pot}) are performed only over the spatial coordinates
while the time coordinate is a fixed parameter.

When in equation (\ref{eq:E coi potenziali}) $\nabla\mathbf{\varphi}\left(\mathbf{r},t\right)=0$,
we can give a physical meaning to the vector potential \textbf{A}.
To do this we have to exchange the roles of the variables \textbf{r}
and \textit{t}; that is, we have to perform an integral over the time
coordinate while the point \textbf{r} remains fixed. Let's consider
a point charge \textit{q }in the position \textbf{r} at a time, which
we will indicate as $-\infty$, when the currents and consequently
the magnetic field are zero. Let's now slowly switch the currents
on. They will generate a magnetic field \textbf{B,} a vector potential
\textbf{A }and therefore an electric field $\mathbf{E}\left(\mathbf{r},t\right)=-\frac{\partial}{\partial t}\mathbf{A}\left(\mathbf{r},t\right)$
that will act on \textit{q.} In order to keep\textit{ q }fixed in\textbf{
r}, an impulse must be applied against the field forces, and this
is given by: 
\begin{equation}
\mathbf{\mathbf{\Upsilon}}\left(\mathbf{r},t\right)=-\int_{-\infty}^{\mathbf{\mathit{t}}}q\mathbf{E}\left(\mathbf{r},t'\right)dt'=\int_{-\infty}^{\mathbf{\mathit{t}}}q\frac{\partial}{\partial t}\mathbf{A}\left(\mathbf{r},t'\right)dt'=q\mathbf{A}\left(\mathbf{r},t\right),\label{eq:impulso}
\end{equation}
where we have adopted the convention of zero vector field $\mathbf{A}$
at the time $t=-\infty.$ The magnetic vector potential can thus be
interpreted as the impulse to be given to the charge \textit{q} to
keep it in a fixed point when the magnetic field rises from zero to
a value \textbf{B}, divided by the charge itself. Magnetic vector
potential can be therefore considered as a ``momemtum vector'',
while the electric scalar potential can be seen as an energy component.

Besides its physical meaning, the magnetic vector potential gives
us also the possibility to write in a clearer and more understandable
way some physical relations. For instance, a mechanical harmonic plane
wave of amplitude $S_{0}$ and angular frequency $\omega$ propagating
in a medium of density $\varrho$ with velocity $v$ carries an intensity
given by:
\begin{equation}
I=\frac{1}{2}\rho\omega^{2}S_{0}^{2}v.\label{eq:intensit=0000E0 1}
\end{equation}
If we consider an electromagnetic linearly polarized, harmonic, plane
wave of amplitude $E_{0}$, angular frequancy $\omega,$ propagating
in a medium of absolute dielectric permittivity $\varepsilon$ with
velocity $v$, its intensity is usually written without explicitly
showing the angular frequency, that is as: 

\begin{equation}
I=\frac{1}{2}\varepsilon E_{0}^{2}v.\label{eq:intensit=0000E0 E}
\end{equation}
The magnetic vector potential gives the possibility to write equation
(\ref{eq:intensit=0000E0 E}) in a form completely similar to equation
(\ref{eq:intensit=0000E0 1}). In fact, from the first term of equation
(\ref{eq:E coi potenziali}) and denoting the vector potential amplitude
with $A_{0}$, we immediately have:

\begin{equation}
I=\frac{1}{2}\varepsilon\omega^{2}A_{0}^{2}v.\label{eq:intensit=0000E0 A}
\end{equation}
Equation (\ref{eq:intensit=0000E0 A}) shows that vector potential
plays for the electromagnetic field the same role played by the displacement
from the equilibrium position for a mechanical wave propagating in
a medium (see equation (\ref{eq:intensit=0000E0 1})). Moreover, since
the intensity, the frequency and the velocity of propagation can be
all measured, equation (\ref{eq:intensit=0000E0 A}) immediately yealds
$A_{0}$ \cite{Giuliani}. 

Equations (\ref{eq:impulso}) and (\ref{eq:intensit=0000E0 A}) and
their interpretations clearly show that the magnetic vector potential
is not a simple mathematical tool, but it has a deep physical meaning
and can greatly help visualization.

\section{Some examples}

In the following we give some examples of calculation of the vector
potential with different and simple strategies, just to show how easy
it can be to visualize vector potential in space (see also \cite{Falstad})
and how problems can be approached from different points of view.
Furthermore, we make some physical considerations that can help highlight
the link between the magnetic vector potential and the electric and
magnetic fields.

\subsection{Magnetic vector potential of a solenoid}

The expression of the magnetic vector potential generated by an infinite
solenoid carrying a current density $\mathbf{J}$ is well known and
can be found in some text-books and many papers. It seems to us that
a very intuitive way of presenting to students this and similar calculations
can be based on the fact that the mathematical relation between $\mathbf{A}$
and $\mathbf{B}$ is the same as that between $\mathbf{B}$ and $\mu_{0}\mathbf{J}$.
In fact, for slowly varying fields, the magnetic field $\mathbf{B}$
is linked to the current density vector $\mathbf{J}$ by the Maxwell
equation:

\begin{equation}
\nabla\times\mathbf{B}=\mu_{0}\mathbf{J},\label{eq:rot B =00003D J}
\end{equation}
while the relation between the magnetic vector potential $\mathbf{A}$
and $\mathbf{B}$ is given by:

\begin{equation}
\nabla\times\mathbf{A}=\mathbf{B\mathrm{.}}\label{eq:rot A =00003D B}
\end{equation}
Moreover, both $\mathbf{A}$ and $\mathbf{B}$ are solenoidal fields.
Therefore, once we know the dependence of $\mathbf{B}$ from $\mathbf{J}$,
we also immediately know the dependence of $\mathbf{A}$ from $\mathbf{B}$
in situations when they have similar symmetry \cite{Semon}. The structure
of equations (\ref{eq:rot B =00003D J}) and (\ref{eq:rot A =00003D B})
could induce to interpret $\mathbf{B}$ as the source of $\mathbf{A}$
(in analogy with the fact that $\mathbf{J}$ is the source of $\mathbf{B}$).
From a didactical point of view we want to stress that this is only
a formal analogy, since the sources of $\mathbf{A}$ are the currents
while the fields can be obtained by deriving $\mathbf{A}$. 

\begin{figure}
\begin{centering}
\includegraphics[scale=0.3]{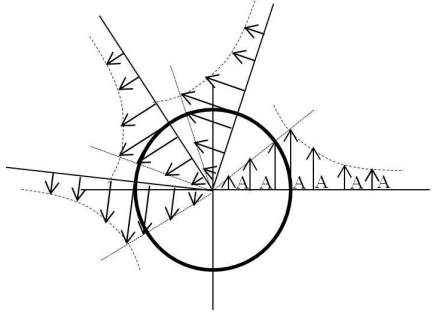}
\par\end{centering}

\label{Figura 1-1}\caption{Field lines of the magnetic vector potential generated by an infinite
solenoid carrying a current.\label{fig: A solenoide}}
\end{figure}

For example, the spatial dependence of $\mathbf{B}$ generated by
an infinite straight wire of radius $a$ is the same as that of $\mathbf{A}$
generated by an infinite solenoid of the same radius, when both wire
and solenoid are carrying a uniform current density. More specifically,
in both cases, the field lines are circular, concentric with the axis
of symmetry and lie on planes perpendicular to this same axis. Therefore,
indicating with $r$ the distance from the symmetry axis, the expressions
of the vector potential inside and outside the solenoid are given
by:
\begin{equation}
A\left(r,t\right)=\frac{1}{2}B\left(t\right)r\;\mathrm{\; and\;\;}A\left(r,t\right)=\frac{1}{2}\frac{a^{2}B\left(t\right)}{r},\:\:\mathrm{respectively},\label{A solenoide}
\end{equation}
see figure \ref{fig: A solenoide}.

If the current, and therefore the magnetic field, is time-dependent,
as in equation (\ref{A solenoide}), an electric field is generated
both inside and outside the solenoid. The expressions of this electric
field can be obtained deriving relations (\ref{A solenoide}):

\begin{equation}
E\left(r,t\right)=-\frac{1}{2}r\frac{d\mathbf{\mathrm{\mathit{B}}}}{dt}\;\mathrm{inside}\;\;\mathrm{and}\;\; E\left(r,t\right)=-\frac{a^{2}}{2}\frac{1}{r}\mathit{\frac{d\mathbf{\mathrm{\mathit{B}}}}{dt}\;\mathrm{outside}}.\label{E solenoide}
\end{equation}

Relations (\ref{E solenoide}) are traditionally obtained from the
integral Maxwell equation $\oint\mathbf{E}\cdot d\mathbf{l}=-\frac{d}{dt}\Phi(\mathbf{B})$.
Following this procedure a strange situation arises because it is
difficult to understand how is it possible that the outside electric
field ``knows'' that, inside the solenoid, $\mathbf{B}$ is changing,
considering that the outside $\mathbf{B}$ is always zero (let's recall
that for slowly varying fields, electromagnetic waves can be neglected).
The question becomes more significant reminding that the field concept
has been introduced just to avoid actions at a distance. The problem
is solved with the introduction of the vector potential, which is
defined both inside and outside the solenoid and it is given by equation
(\ref{A solenoide}). From this equation, using equation (\ref{eq:E coi potenziali})
and keeping in mind that no free charges are present, one can obtain
the electric field. Thus, it is the local time dependent $\mathbf{A}$
that generates $\mathbf{E}$. 

We note that the use of the local form $\nabla\times\mathbf{E=-\mathrm{\nicefrac{\partial\mathbf{B}}{\partial t}}}$
tells us only that outside the solenoid the electric field is irrotational
(even if it is obviously not conservative since the region is not
simply connected) and does not give the explicit expression for the
electric field.

\subsection{Magnetic vector potential of two parallel planes}

\begin{figure}
\begin{centering}
\includegraphics[scale=0.3]{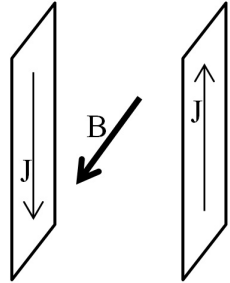}
\par\end{centering}

\label{Figura 1}\caption{Magnetic field generated by two planes carrying antiparallel currents.\label{fig:piani paralleli}}
\end{figure}

\begin{figure}
\begin{centering}
\includegraphics[scale=0.3]{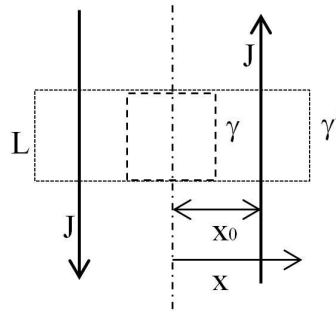}
\par\end{centering}

\label{Figura 2}\caption{The vertical arrows represent the section of the planes of the currents;
$\gamma$ e $\gamma'$ are the boundaries of the rectangles chosen
to calculate the circulation of $\mathbf{A}$. \label{fig:rettangoli piani}}
\end{figure}

Let's now consider two parallel planes carrying, in opposite directions,
a uniform current of density $\mathbf{J}$ per unit length. It is
straightforward to understand that between the planes there is a uniform
magnetic field $\mathrm{\mathbf{B}}$, while outside the planes the
magnetic field is zero (see figure \ref{fig:piani paralleli}). To
determine the vector potential we could integrate equation (\ref{eq:A in funzione di J})
or, in a simpler way, we can again start from equation (\ref{eq:A in funzione di J}),
but now just to understand the symmetries of $\mathbf{A}$ in terms
of those of $\mathbf{J}$, to later obtain $\mathbf{A}$, by solving
equation (\ref{eq:circuitazione}). From figure \ref{fig:piani paralleli}
it follows immediately that the magnetic vector potential is parallel
to the currents in the planes. Therefore we can choose as the surface
$\Sigma$ of equation (\ref{eq:circuitazione}) a rectangle lying
in a plane normal to the planes of the currents and with symmetry
axis in the median plane, as shown in figure \ref{fig:rettangoli piani}. 

\begin{figure}
\begin{centering}
\includegraphics[scale=0.3]{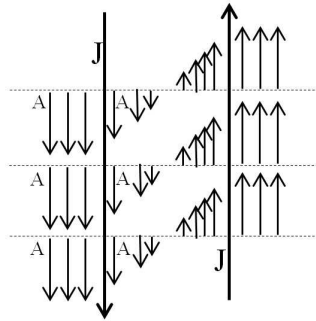}
\par\end{centering}

\label{Figura 3}\caption{Field lines of the magnetic vector potential generated by two planes
carrying opposite currents.\label{figura A piani}}
\end{figure}

With the symbology of figure \ref{fig:rettangoli piani}, from equation
(\ref{eq:circuitazione}) we get:

\begin{equation}
A\left(x\right)=Bx\text{,}\qquad\mathrm{between\: the\: planes}\label{A interno piani}
\end{equation}
and

\begin{equation}
A\left(x\right)=Bx_{0}\qquad\mathrm{outside\: the\: planes}.\label{piani paralleli tre-1-1}
\end{equation}
The field lines of $\mathbf{A}$ are given in figure \ref{figura A piani}.
It is interesting to observe that, as in the previous case, the vector
potential is different from zero both inside and outside the planes. 

\subsection{Thoughts on the link between the magnetic field and the magnetic
vector potential }

In the previous examples we found the expression of the vector potential
starting from a known current distribution. It can be interesting
now to determine $\mathbf{A}$ through equation (\ref{B =00003D rot A}),
that is, starting from a known magnetic field $\mathbf{B}$. We already
know that the problem is not univocally determined. However, what
discussed in the previous section allows us to shed some light on
the physical implications of this fact.

Let's imagine to calculate $\mathbf{A}$ for a uniform $\mathrm{\mathbf{B}}.$
When we choose a particular class of close paths to calculate the
circulation of the vector potential, the symmetry of the problem is
broken and a particular $\mathbf{A}$ is found. To recover the lost
physical symmetry, one generally considers equivalent all the vector
potentials generating the same field $\mathrm{\mathbf{B}};$ and in
a sense this is one of the physical meaning of the gauge invariance.
Back to our example, if $\mathrm{\mathbf{B}}$ is really uniform in
the whole space we don't know whether we are inside an infinite solenoid
of infinite radius or between a couple of current carrying planes,
infinite distance apart. Therefore, even if we are in the same Coulomb
gauge (in our appoximation the gauge is fixed), $\mathbf{A}$ is not
univocally determined by $\mathrm{\mathbf{B}}$ because, as we have
already said, the currents which could generate this field do not
vanish at infinity. It is clear that the currents determine both $\mathbf{B}$
and $\mathbf{A}$; the potential $\mathbf{A}$ determines $\mathbf{B}$,
while the viceversa is not true.

\section{Conclusions}

Two main facts hinder the comprehension and therefore the use of the
magnetic vector potential. The first one is the non univocity of $\mathbf{A}$
implied by its definition given by equation (\ref{B =00003D rot A});
the second one is the scarce discussion traditionally devoted to its
physical meaning.

Convinced of the educational value of the vector potential in dealing
with many physical situations, we have developed a path which in our
opinion can overcome the above stated difficulties. Starting from
the generalized Ampère-Laplace low, we attained an expression of $\mathbf{A}$
in terms of its empirical referent, i. e. the conduction current density,
for slowly time-dependent electric and magnetic fields. Traditionally,
this result is obtained working with static fields or starting from
the wave equation for $\mathbf{A}$. Our approach has the advantage
of being much more general than that with static fields, principally
because within our quasi-static approximation we can clearly give
a physical meaning to $\mathbf{A}$. Moreover we believe that our
approach can be presented in a basic course on electromagnetism before
the study of the electromagnetic wave equations, thus giving students
the time to familiarize with the concept. In our work we found a privileged
gauge (the Coulomb gauge) and the physical meaning of $\mathbf{A}$
was discussed in a similar way of that of the electric scalar potential,
another fact which can help comprehension. In addition, in some circumstances,
the use of the vector potential allowed us a causal local description
clearer than that given by the magnetic field alone and to highlight
interesting parallelisms with mechanical situations. To conclude,
we firmly consider the introduction of the magnetic vector potential
in electromagnetism not only a good tool for making calculations,
but also a useful way to better understand many physical phenomena.

\ack{}{}

The authors are grateful to Tommaso Maccacaro and Claudio Fazio for
a careful reading of the paper and useful suggestions.

\end{document}